\documentclass[letterpaper]{article}
\usepackage{emulateapj}
\usepackage{float,epsfig,psfig}
\begin{document}

\title{
DISCOVERY OF SOFT X-RAY EMISSION FROM IO, EUROPA AND THE IO PLASMA TORUS}

\author{
Ronald F. Elsner\altaffilmark{1}, 
G. Randall Gladstone\altaffilmark{2}, 
J. Hunter Waite\altaffilmark{3},
Frank J. Crary\altaffilmark{3},
Robert R. Howell\altaffilmark{4},
Robert E. Johnson\altaffilmark{5},
Peter G. Ford\altaffilmark{6},
Albert E. Metzger\altaffilmark{7},
Kevin C. Hurley\altaffilmark{8},
Eric D. Feigelson\altaffilmark{9},
Gordon P. Garmire\altaffilmark{9},
Anil Bhardwaj\altaffilmark{10},
Denis C. Grodent\altaffilmark{3}, 
Tariq Majeed\altaffilmark{2}, 
Allyn F. Tennant\altaffilmark{1}, 
Martin C. Weisskopf\altaffilmark{1} 
} 

\altaffiltext{1}{Space Science Department, NASA Marshall Space Flight Center, SD50, Huntsville, AL 35812} 
\altaffiltext{2}{Department of Space Science, Southwest Research Institute, P.O. Drawer 28510, San Antonio, TX 78228}
\altaffiltext{3}{Department of Atmospheric, Oceanic and Space Sciences, University of Michigan, Ann Arbor, MI 48109}
\altaffiltext{4}{Department of Physics and Astronomy, University of Wyoming, P.O. Box 3905, University Station, Laramie, WY 82071}
\altaffiltext{5}{Department of Engineering Physics, Thornton Hall, University of Virginia, Charlottesville, VA 22903}
\altaffiltext{6}{Center for Space Research, Massachusetts Institute of Technology, Cambridge, MA 02139}
\altaffiltext{7}{Jet Propulsion Laboratory, Pasadena, CA 91109}
\altaffiltext{8}{Space Science Laboratoy, University of California at Berkeley, Berkeley, CA 94720}
\altaffiltext{9}{Department of Astronomy and Astrophysics, 525 Davey Laboratory, Pennsylvania State University, State College, PA 16802}
\altaffiltext{10}{Space Physics Laboratory, Vikram Sarabhai Space Centre, Trivandrum, India}

\begin{abstract}

We report the discovery of soft (0.25--2 keV) x-ray emission from the Galilean satellites Io and Europa, probably Ganymede, and from the Io Plasma Torus (IPT).  
Bombardment by energetic ($>$ 10 keV) H, O, and S ions from the region of the IPT seems the likely source of the x-ray emission from the Galilean satellites.
According to our estimates, fluorescent x-ray emission excited by solar x-rays, even during flares from the active Sun, charge-exchange processes, previously invoked to explain Jupiter's x-ray aurora and cometary x-ray emission, and ion stripping by dust grains fail to account for the observed emission.   
On the other hand, bremsstrahlung emission of soft X-rays from non-thermal electrons in the few hundred to few thousand eV range may account for a substantial fraction of the observed x-ray flux from the IPT.

\end{abstract}

\keywords{X rays: general --- planets and satellites: individual (Io, Europa, Jupiter)}

\section{Introduction} \label{s:introduction}

Imaging and spectral data from the infrared through the extreme ultraviolet provide important information on the makeup of the surfaces and atmospheres of the Galilean satellites Io, Europa, Ganymede, and Callisto (Carlson {\sl et al.} 1999, Barth {\sl et al.} 1997, Hall {\sl et al.} 1998), discovered by Galileo Galilei in 1610, and of the Io plasma torus (IPT; Hall {\sl et al.} 1994, Woodward, Scherb, \& Roesler 1997, Gladstone \& Hall 1998).  
Here we report the results of high-spatial resolution x-ray observations of the Jovian system with the {\sl Chandra X-ray Observatory} showing that the Galilean satellites and the IPT are also x-ray emitters, and we discuss the physical processes that may contribute.

\section{Observations} \label{s:observations}

The {\sl Chandra X-ray Observatory} (Weisskopf {\sl et al.} 2000) observed the Jovian system on 25-26 Nov 1999 with the Advanced CCD Imaging Spectrometer (ACIS), in support of the Galileo flyby of Io, and on 18 Dec 2000 with the imaging array of the High Resolution Camera (HRC-I), in support of the Cassini flyby of Jupiter (Gladstone {\sl et al.} 2001). 
During the ACIS observations, spanning 86.4 ks, the telescope focus and the planet were placed on the back-side illuminated CCD designated S3 in order to take advantage of this CCD's sensitivity to low energy x-rays.  
The spacecraft was repointed four times to allow for the planet's motion and was always oriented to allow the planet to move along the CCD's second node.  
These data were corrected for charge-transfer-inefficiency effects (Townsley {\sl et al.} 2000), and only the standard ASCA grades 0, 2, 3, 4, and 6 were retained in order to reduce the background induced by charged particles.  
Charge-transfer-inefficiency refers to the loss of charge during CCD readout for pixels furthest away from the readout node, leading to a low estimate for the event energy.
Event grades depend on the distribution of charge induced by the event about the center pixel.  
X-ray and charge particle events produce different event grade distributions, allowing a powerful way to reduce unwanted background events.
One ACIS pixel is 0.492 arcsec by 0.492 arcsec and the half-power diameter of the {\sl Chandra} system level point-spread function is about 0.8 arcsec.
No repointings were necessary during the HRC-I observations, spanning 36.0 ks (approximately one rotation of Jupiter), because of the instrument's larger field of view and the observation's shorter length.  
One HRC-I pixel is 0.1318 arcsec by 0.1318 arcsec.

Although the ACIS S3 optical filter is optically thick to optical light, there is an interference peak in its transmission near $\sim$9,000\AA \ ($\sim$1.4 eV) which is likely to affect ACIS S3 observations of solar system objects and nearby bright late-type stars.
Indeed, the ACIS S3 data for Jupiter were compromised by the throughput in low energy channels of large numbers of optical photons from the planet's bright disk.
To eliminate this problem for our analysis, all events within 34.4 arcsec ($\sim$1.4 times Jupiter's radius) of the planet's center were removed from the data. 
In any event, removal of this data is required for analysis of the Galilean satellites in order to avoid contamination by events associated with Jupiter.
The planet's optical emission also produced a bump in the ACIS bias frame, and all events within 34.4 arcsec of the center of this bump were removed from the data.
For sufficiently bright x-ray sources, x-ray events occurring during the 41 ms necessary to read out each 3.2 s ACIS accumulation appear as a streak, or trailed image, in the data.
For our ACIS Jupiter observations, optical photons from the planet's disk do not contribute to a trailed image, as 41 ms is too short a time for sufficient accumulation to mimic an x-ray event.
The trailed image from Jupiter's true x-ray events is a small effect for 
E $>$ $\sim$440 eV, where we have reasonably accurate knowledge of Jupiter's x-ray spectrum.  
From an extrapolation of Jupiter's x-ray spectrum to lower energies, we infer that the trailed image is also not a dominant effect for our analysis of the Io Plasma Torus at energies down to 250 eV.
The trailed image is unimportant for our analysis of x-ray emission from the Galilean satellites.
Since we wished to remove x-ray events associated with Jupiter, we removed events within 34.4 arcsec of the planet's center from the HRC-I data.

\subsection{The Galilean satellites} \label{ss:moons}

\begin {table*} [htb] 
\caption {X-ray emission from the Galilean satellites}
\label{tab:moons}
\begin {center}
\begin {tabular}{cccccc} 
\hline 
\hline
               &             & Io & Europa & Ganymede & Callisto \\ 
\hline
   ACIS$^a$            & $N_{moon}^{b}$  & 11 &  12 & 5 & 3 \\ 
 (Nov 25-26, 1999) & $<N_{bkgd}>^{c}$ & 1.26 & 1.94 & 1.25 & 1.27 \\ 
     & $Pr[N \ge N_{moon}]^{d}$  & 1.03 10$^{-8}$ & 1.46 10$^{-7}$ & 9.09 10$^{-3}$ & 0.137 \\ 
\hline 
     HRC-I          & $N_{moon}^{b}$  & 10 & 3 & 3 & 6 \\ 
    (Dec 18, 2000)        & $<N_{bkgd}>^{e}$ & 2.88 & 3.08 & 3.58 & 3.24 \\ 
     & $Pr[N \ge N_{moon}]^{d}$  & 8.06 10$^{-4}$ & 0.595 & 0.694 & 0.110 \\ 
\hline 
Joint (ACIS+HRC-I)  & $Pr[N \ge N_{moon}]^{d}$  &  3.47 10$^{-9}$ & 2.36 10$^{-4}$ & 0.116 & 0.041 \\ 
\hline 
\end {tabular}
\end {center}
${}^a$ 250--2,000 eV. \\ 
${}^b$ Number of counts in detect cell centered on the satellite. \\ 
${}^c$ Average number of counts per detect cell in a 98.9 by 98.9 arcsec region surrounding the satellite. \\ 
${}^d$ Probability of chance occurence for $N \ge N_{moon}$. \\ 
${}^e$ Average number of counts per detect cell in a 98.7 by 98.7 arcsec region surrounding the satellite.
\end {table*}

\begin {figure*} [htb]
\centerline {\epsfxsize=0.90\hsize \epsffile{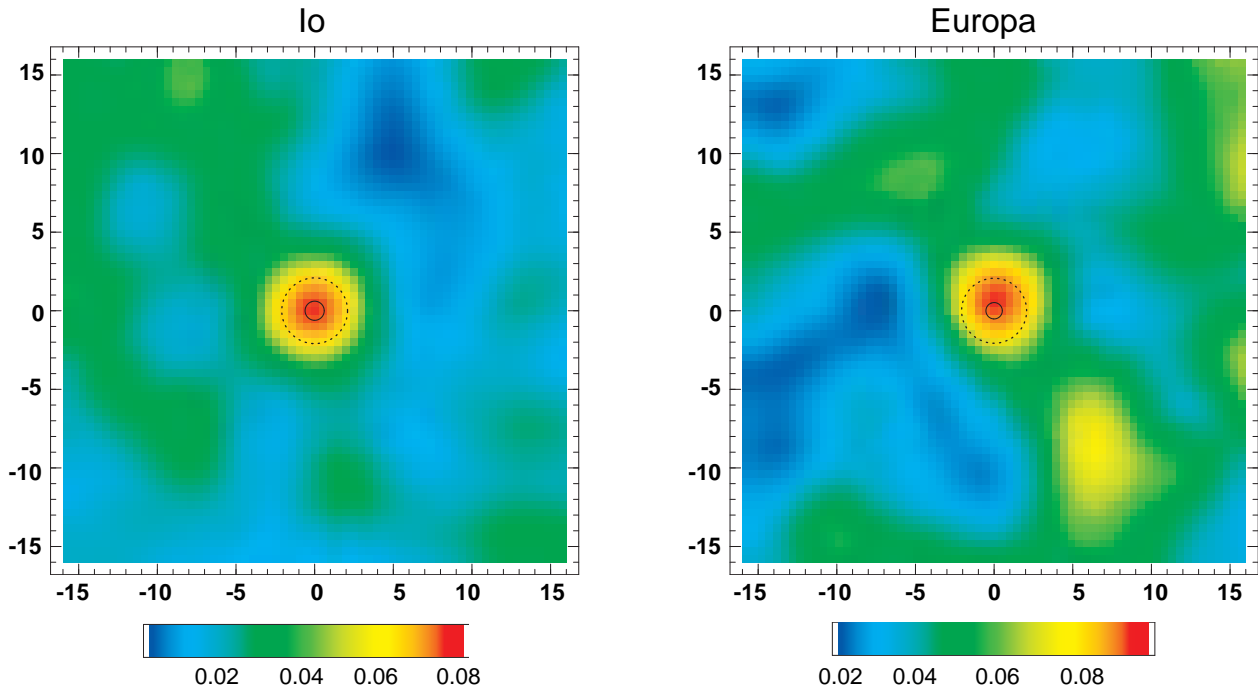} }
\caption {\label{fig:ioeuropa} 
ACIS images of Io and Europa (250 eV $<$ E $<$ 2,000 eV).  
The images have been smoothed by a two dimensional gaussian with $\sigma$ = 2.46 arcsec (5 ACIS pixels).  
The axes are labelled in arcsec (1 arcsec $\simeq$ 2995 km) and the scale bar is in units of smoothed counts per image pixel (0.492 by 0.492 arcsec).
The solid circle shows the size of the satellite (the radii of Io and Europa are 1821 km and 1560 km, respectively), and the dotted circle the size of the detect cell.
}
\end{figure*}

In order to search for emission from each Galilean satellite, we removed {\sl Chandra's} orbital motion and transformed the data into a co-moving frame using an ephemeris obtained from the Jet Propulsion Laboratory.
The size of the detect cells was set equal to the root-sum-square of the satellite's radius and 2 arcsec.
For the background, we determined an average number of events per detect cell in a larger region around the satellite in its own frame.
This region was 98.9 by 98.9 arcsec for ACIS and 98.7 by 98.7 arcsec for HRC-I.  
The number of events within the detect cell centered on each satellite compared with this background value determines a probability of chance occurrence according to Poisson statistics.
The results of this analysis appear in Table~\ref{tab:moons}.
Since these objects move across the field of view, {\sl Chandra's} spatial resolution at their position varies, so in each case we also compared the radial distribution of events around the satellite with that around 100 points chosen at random in the surrounding region.
Io is detected in both the ACIS and HRC-I data, while Europa is detected in the ACIS data only.  
Ganymede appears at the 99\% confidence level in the ACIS data.
A simple calculation of joint probabilities confirms the detection of Io and Europa, with results that we find tantalizing for Ganymede and Callisto.
The nominal energies of the x-ray events range from 300 eV to 1890 eV and seem to show a clustering between 500 and 700 eV.
The mean energy, weighted appropriately by {\sl Chandra's} effective area, of the events are 515 eV and 664 eV, for Io and Europa, respectively.  
The corresponding estimated energy flux at the telescope and emitted power from the satellite are $4.1 \times 10^{-16}$ erg/s-cm$^2$ and 2.0 MW for Io, and $3.0 \times 10^{-16}$ erg/s-cm$^2$ and 1.5 MW for Europa. 
Figure~\ref{fig:ioeuropa} shows smoothed ACIS images of Io and Europa.  

\subsection{The Io Plasma Torus} \label{ss:torus}

\begin {figure*} [htb]
\centerline {\epsfxsize=0.90\hsize \epsffile{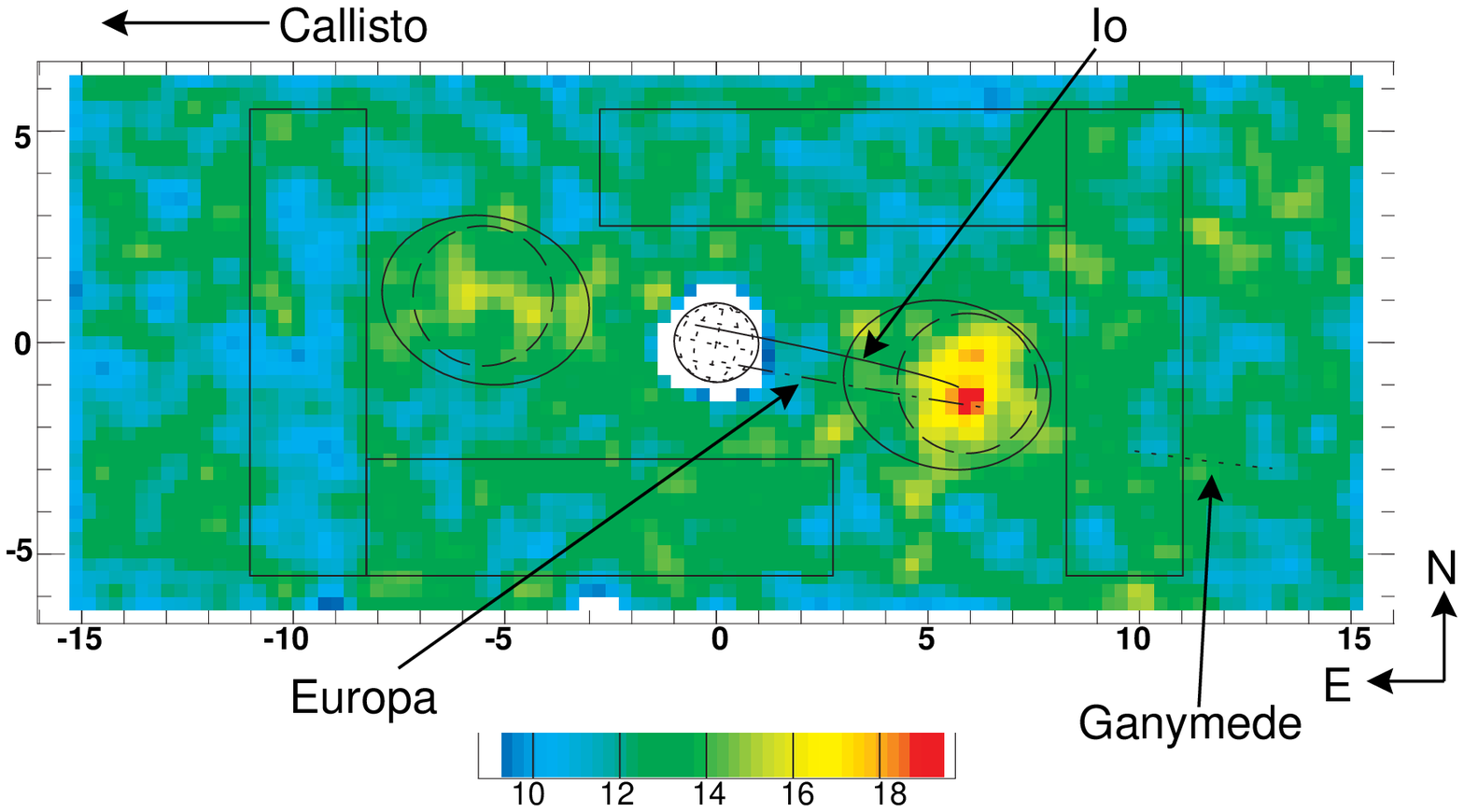} }
\caption {\label{fig:ipthrc} 
HRC-I image of the IPT (Dec 18, 2000).  
The image has been smoothed by a two dimensional gaussian with $\sigma$ = 7.38 arcsec (56 HRC-I pixels).  
The axes are labelled in units of Jupiter's radius, R$_{J}$, and the scale bar is in units of smoothed counts per image pixel (7.38 by 7.38 arcsec).
The paths traced by Io (solid line, semi-major axis 5.9 R$_{J}$), Europa (dashed line, semi-major axis 9.5 R$_{J}$), and Ganymede (dotted line, semi-major axis 15.1 R$_{J}$) are marked on the image.
Callisto (semi-major axis 26.6 R$_{J}$) is off the image to the dawn side, although the satellite did fall within the full microchannel plate field of view.
For this observation, Jupiter's equatorial radius corresponds to 23.9 arcsec.
The regions bounded by rectangles were used to determine background.
The regions bounded by dashed circles or solid ellipses were defined as source regions.
}
\end{figure*}

Figure~\ref{fig:ipthrc} shows the HRC-I image zoomed back from Jupiter, in the planet's frame of reference.
There is a region of diffuse emission on the dusk side of the planet, and a fainter region of diffuse emission on the dawn side of the planet.
This morphology is similar to the dawn-dusk asymmetry seen in EUVE data (Hall {\sl et al.} 1994, Gladstone \& Hall 1998).
The paths traced by Io, Europa, and Ganymede during the HRC-I observation are also shown.
Io and Europa pass through the dusk side region of diffuse x-ray emission.
We associate these regions of diffuse emission with the IPT.
The background subtracted HRC-I count rates are $3.2 \times 10^{-3}$ counts/s-arcmin$^{2}$ on the dawn side of Jupiter and $5.6-7.1 \times 10^{-3}$ counts/s-arcmin$^{2}$ on the dusk side of Jupiter.
The larger rate for the dusk side region applies to a circular source region of 1.36 arcmin$^{2}$, and the smaller to the larger elliptical source region of 2.43 arcmin$^{2}$ suggested by the ACIS observations.
The rate for the dawn side region applies to both the smaller and larger source regions.

\begin {figure*} [htb]
\centerline {\epsfxsize=0.90\hsize \epsffile{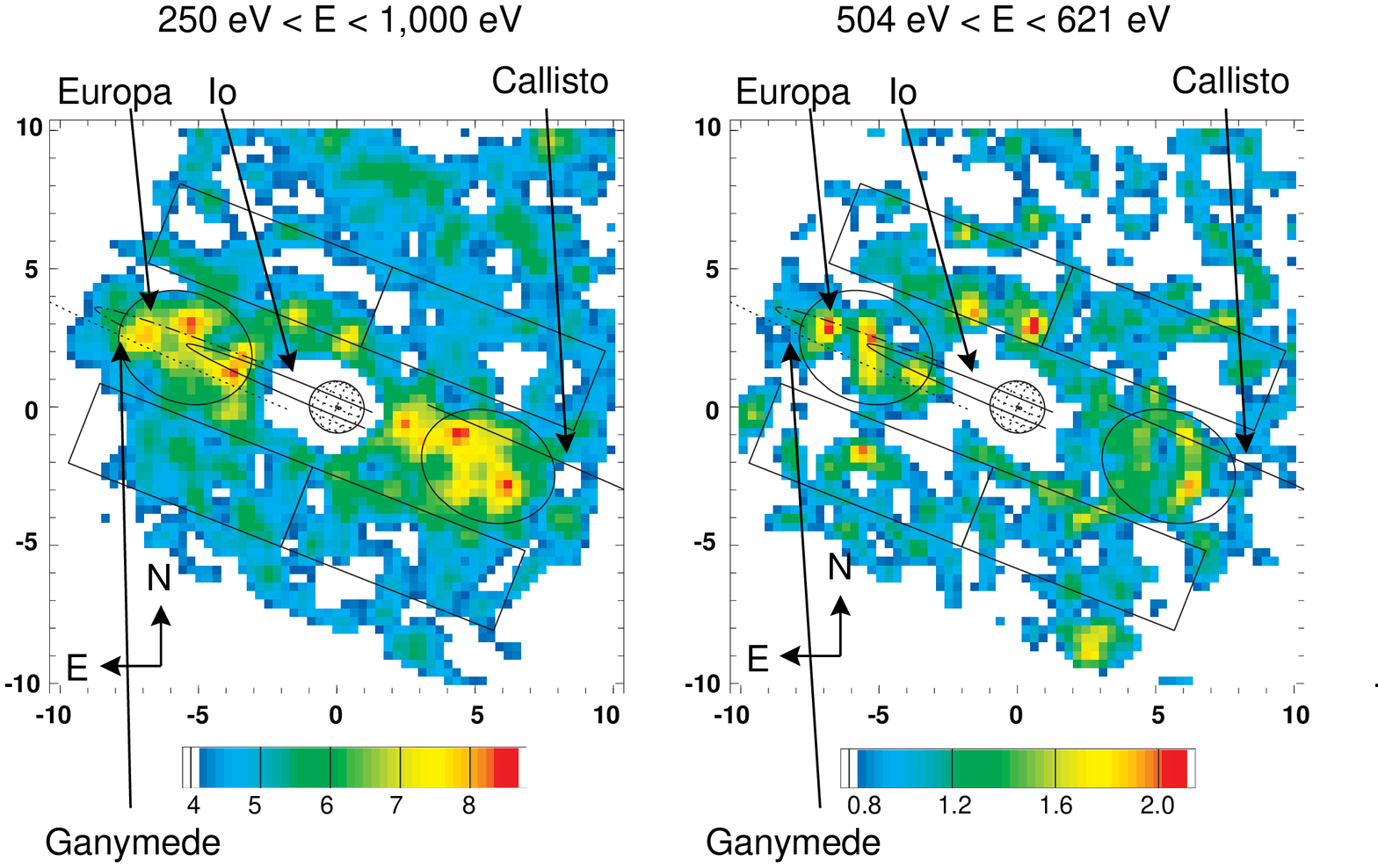} }
\caption {\label{fig:iptacis} 
ACIS image of the IPT (Nov 25-26, 1999) for 250 eV $<$ E $<$ 1,000 eV (left) and for 504 $<$ E $<$ 621 eV (center).  
The image has been smoothed by a two dimensional gaussian with $\sigma$ = 7.38 arcsec (15 ACIS pixels).  
The axes are labelled in units of Jupiter's radius, R$_{J}$ and the scale bar is in units of smoothed counts per image pixel (7.38 by 7.38 arcsec).
For this observation, Jupiter's equatorial radius corresponds to 23.8 arcsec.
The paths traced by Io (solid line on the dawn side), Europa (dashed line), Ganymede (dotted line), and Callisto (solid line on the dusk side) are marked on the image.
The regions bounded by rectangles were used to determine background.
The regions bounded by ellipses were defined as source regions.
The exposure (right) varies over the image because four overlapping observations were concatenated, and 
because we discarded data within 34.4 arcsec of Jupiter's center or within 34.4 arcsec of the center of the bump in the bias frame.
}
\end{figure*}

Figure~\ref{fig:iptacis} shows the ACIS image zoomed back from Jupiter, in the planet's frame of reference over the energy ranges 250--1,000 eV (left panel) and 504--621 eV (center panel).
Because this image was produced from four overlapping observations, and also because of the instrumental effects discussed above, we  show an exposure map (right panel) for the image in Figure~\ref{fig:iptacis}.  
The area of low exposure around Jupiter is due to the removal of events within 34.4 arcsec of the planet's center or within 34.4 arcsec of the center of the bump in the bias frame.
The ACIS image also shows regions of diffuse emission on the dawn side and dusk side of the planet.
The paths traced by Io, Europa, Ganymede, and Callisto during the ACIS observations are also shown.
In this case Io and Europa pass through the dawn side region of diffuse x-ray emission.
The ACIS 250--1,000 eV background subtracted count rates are $1.7 \times 10^{-3}$ counts/s-arcmin$^{2}$ both on the dawn side and dusk side of Jupiter.
However, the 504--621 eV image shows significant asymmetry with stronger emission on the dawn side where Io and Europa were located during the ACIS observations.
The HRC-I image also shows stronger emission on the side (dusk in that observation) where Io and Europa were located during that observation.
The narrow band ACIS image also shows emission to the north of Jupiter, which we do not associate with the IPT.

\begin {figure*} [htb]
\centerline {\epsfxsize=0.90\hsize \epsffile{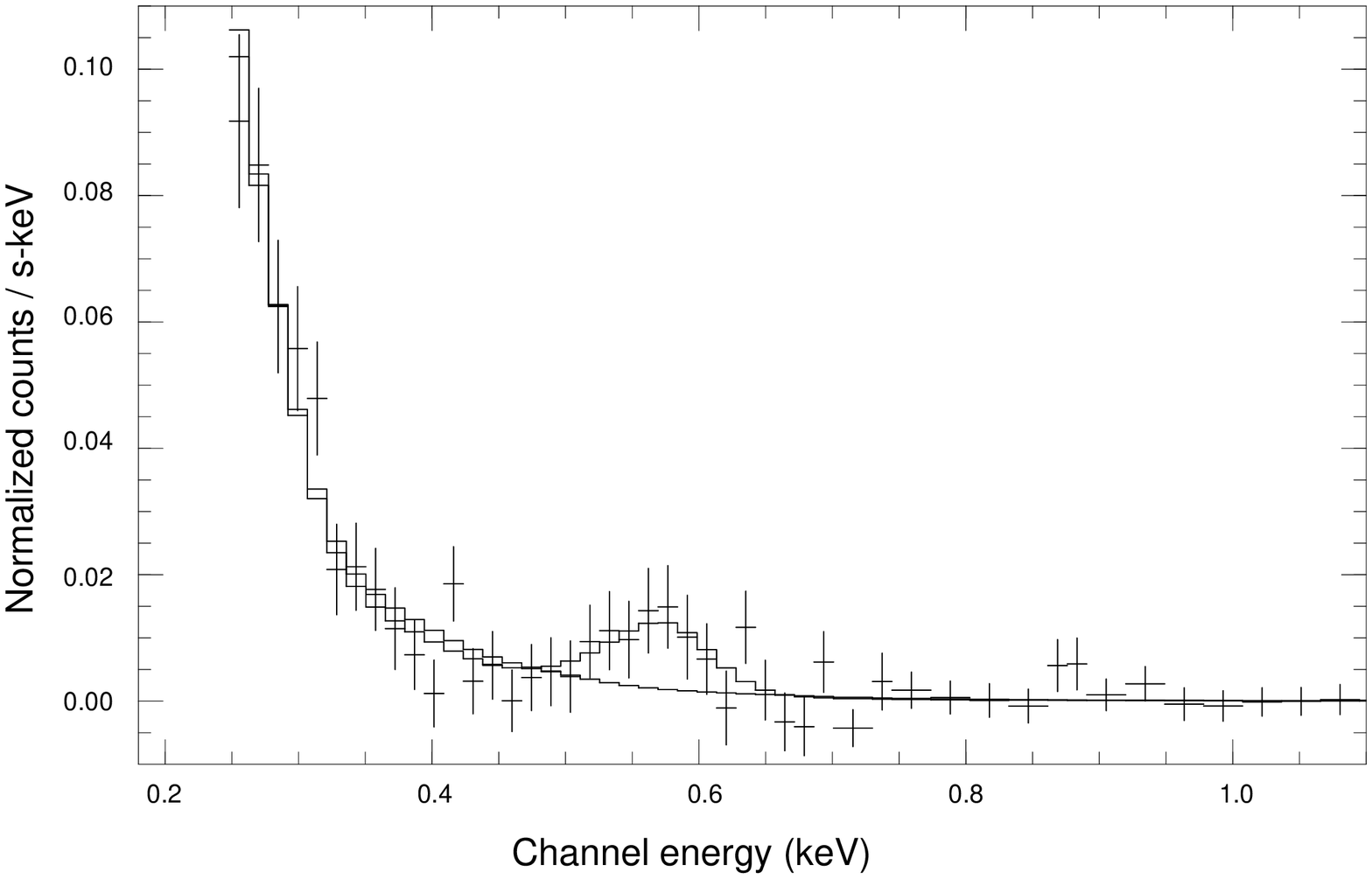} }
\caption {\label{fig:iptspec} 
The background subtracted ACIS IPT spectrum, normalized counts/s-keV vs channel energy in keV, for 0.25 keV $<$ E $<$ 1.0 keV, together with the power-law and power-law plus gaussian model fits given in Table~\ref{tab:ipt}.
}
\end{figure*}

Both a simple power-law and thermal bremsstrahlung give statistically acceptable fits to the background subtracted 250--1,000 eV ACIS IPT spectrum (see Table~\ref{tab:ipt} and Figure~\ref{fig:iptspec}).
However, adding a single gaussian line does significantly improve these fits.  
The central energy of the gaussian is very close to a strong O VII line at 574 eV.
Lines from other oxygen charge states may also contribute.  
The background subtraction procedure correctly removed instrumental lines such as Si K$\alpha$ at 1.74 keV.  
The 250--1,000 eV energy flux at the telescope from the IPT (within the ovals marked on Figure~\ref{fig:iptacis}) is $ 2.4 \times 10^{-14}$ erg/s-cm$^{2}$, corresponding to a luminosity of 0.12 GW, and is approximately evenly divided between the dawn side and dusk side.  
The apparent x-ray line emission around 570 eV originates predominantly on the dawn side of the planet.
EUVE observations in 1993 (Hall {\sl et al.} 1994, Gladstone \& Hall 1998) determined an IPT energy flux of $ (7.2 \pm 0.2) \times 10^{-11}$ erg/s-cm$^{2}$ over the interval 370--735 \AA \ (17--34 eV).
This flux is $\sim$3,000 times that inferred from the {\sl Chandra} data over the interval 250--1,000 eV.

\begin {table*} [htb]
\caption {Spectral fits to ACIS IPT data (250 eV $<$ E $<$ 1,000 eV)}
\label{tab:ipt}
\begin {center}
\begin {tabular}{ccccc} 
\hline 
\hline
\multicolumn{5}{c}{Power-law:  $dN/dE = K E^{-\gamma}$} \\
\multicolumn{5}{c}{$\chi^{2} = 48$ for 51 degrees of freedom} \\ 
\hline
 & \multicolumn{2}{c}{ $\gamma$ } & \multicolumn{2}{c}{ K$^{a}$ }   \\ 
\hline
 & \multicolumn{2}{c}{6.3(+0.4/-0.4)} & \multicolumn{2}{c}{1.5(+1.8/-0.9)} \\ 
\hline 
\hline
\multicolumn{5}{c}{Power-law plus gaussian line:
$dN/dE = K E^{-\gamma} + (A/\sigma \sqrt{2 \pi}) exp[-(E-E_{line})^{2}/2 \sigma^{2}]$} \\
\multicolumn{5}{c}{$\chi^{2} = 34$ for 48 degrees of freedom} \\ 
\hline
 $\gamma$ & K$^{a}$  & E$_{line}$ (eV) & $\sigma$ (eV) & A$^{d}$  \\ 
\hline
    6.8(+0.5/-0.4)     &   0.8(+0.6/-0.4)      &  569(+11/-13)    &   0(+25) &    3(+1/-1)     \\
\hline 
\hline
\multicolumn{5}{c}{Thermal bremsstrahlung: $dN/dE = EM f_{brems}(E,T)$} \\
\multicolumn{5}{c}{$\chi^{2} = 50$ for 51 degrees of freedom} \\ 
\hline
   & \multicolumn{2}{c}{ T$^{b}$ } & \multicolumn{2}{c}{EM${}^{c}$/10$^{40}$}   \\ 
\hline
 & \multicolumn{2}{c}{60(+7/-5)} & \multicolumn{2}{c}{4.9(+2.4/-1.9)} \\ 
\hline 
\hline
\multicolumn{5}{c}{Thermal brems plus gaussian line: 
$dN/dE = EM f_{brems}(E,T) + (A/\sigma \sqrt{2 \pi}) exp[-(E-E_{line})^{2}/2 \sigma^{2}]$} \\
\multicolumn{5}{c}{$\chi^{2} = 31$ for 48 degrees of freedom} \\ 
\hline
 T$^{b}$ & EM${}^{c}$/10$^{40}$  & E$_{line}$ (eV) & $\sigma$ (eV) & A$^{d}$  \\ 
\hline
  56(+6/-5)     &   7(+4/-2)      &  567(+10/-12)    &   0(+28) &    4(+1/-1)     \\
\hline
\end {tabular}
\end {center}
${}^a$ In units of 10$^{-10}$ counts/s-eV-cm$^{2}$. \\
${}^b$ In units of eV. \\
${}^c$ $\int n_{e} n_{i} dV$, with $n_{e}$ and $n_{i}$ the electron and ion densities in units of cm$^{-3}$. \\
${}^d$ In units of 10$^{-6}$ counts/s-cm$^{2}$. \\
\end {table*}

\section{Discussion} \label{s:discussion}

We have observed x-rays that originate from the IPT, the Galilean satellites Io and Europa, and probably Ganymede. 
Two known energy sources are available to drive these processes: (1) the solar x-ray flux; and (2) the energetic particle populations in the region of the IPT. 
Furthermore, we must consider the x-ray production processes that originate from the surfaces of the Galilean satellites as distinct and different from the x-ray processes that lead to emission from the IPT. 

\subsection{The Galilean satellites} \label{ss:dismoons}

We first consider the surface emissions from Io and Europa.
Based on the results of model calculations (Peres {\sl et al.} 2000), the energetic particles that constitute the outer radiation belts of Jupiter, and are found within the IPT, provide energy fluxes to the surfaces of Io and Europa  $\sim$20 and $\sim$110, respectively, times larger than the solar x-ray flux during solar maximum conditions with active flares.
We show below that the relevant incident particles and energies are those which deposit most of their energy in the top $\sim$10 micron of the surface.  
Irradiation models show that ion energy losses dominate that from electrons in this layer (Paranicas, Carlson, \& Johnson 2001, Cooper et al. 2001).
Even along Europa's trailing hemisphere, we suspect the total energetic ion
dose dominates the energetic electron dose down to about tens of microns in
depth, below which point the situation is reversed (Paranicas, Carlson, \& Johnson 2001).  
This, together with the clustering of x-ray event energies between 500 and 700 eV, suggests that the observed emission cannot be due to electron bremsstralung.
Since the  cross-sections for interaction and the efficiency for x-ray production from energetic ion impact and from fluorescence excited by solar x-rays are comparable, we conclude that the x-ray emission from Io and Europa is ultimately powered by the energetic ion population in the IPT.
The surfaces of the Galilean satellites are exposed to bombardment by energetic ions diffusing radially inward through the region of the IPT.
The incident flux is made up of energetic H, O and S ions (Paranicas {\sl et al.} 2001, Cooper {\sl et al.} 2001).  
Although Io and Europa have ionospheres which deflect low energy ions, more energetic ions with large gyro-radii will reach their surfaces and produce x-rays through particle-induced x-ray emission (Johansson, Campbell, \& Malmqvist 1995).
In fact the large gyro-radii for the x-ray producing ions means the effective cross-section of the Galilean satellites to IPT ions is larger than their physical size (Pospieszalska \& Johnson 1989).

Io's atmosphere arises from sublimation of surface SO$_{2}$ frost, sputtering of the surface by Jovian magnetospheric particles, and volcanic activity, with the latter perhaps being the dominant contributor.
This atmosphere is patchy rather than uniform, but can be expected to provide a lower limit to the energy of ions impacting the surface.  
Approximately 5 keV H$^+$, 40 keV O$^+$, and 60 keV S$^+$ ions penetrate an atmosphere with a column density of $10^{17}$ SO$_{2}$ molecules/cm$^{2}$ (Lanzerotti et al. 1982); these limits fall between 1 and 10 keV for a column density of $10^{16}$ SO$_{2}$ molecules/cm$^{2}$.  
For comparison, electrons with energy greater than about 10 keV penetrate a column density of $10^{17}$ SO$_{2}$ molecules/cm$^{2}$ (Michael and Bhardwaj 2000).
The integrated ion energy flux at Europa for H, O, and S above 10 keV (Paranicas {\sl et al.} 2001, Cooper {\sl et al.} 2001) is $\sim 10^{10}$ keV/s-cm$^{2}$.
The energetic ion flux at Io is $\sim$5 times smaller, but is less well known than for Europa.
The depth of the region of interest is limited by the absorption of escaping x-rays in the surface.  
For Europa's water ice surface, absorption in oxygen dominates, while for Io's  surface, absorption in oxygen is complemented by that in silicon, sulfur, and other elements.
Oxygen K$\alpha$ x-rays at 525 eV in ice have an optical depth of unity at about 9.5 micron, implying an oxygen column density of 3.2 $\times$ 10$^{19}$ cm$^{-2}$.
For incident protons with energy $>$ $\sim$300 keV, and an oxygen target, the cross-section for x-ray production is flat with energy and is $\sim10^{-21}$ cm$^{2}$, implying an x-ray yield per proton of $\sim$0.01 for Europa.
Similarly, the presence of sulfur dioxide (atomic weight 64) rather than water (atomic weight 18) suggests an an x-ray yield per proton $\sim$0.04 for Io.  
These yield estimates are uncertain; accurate estimates require more detailed work.
The production cross-sections for energetic O and S ions are larger but the corresponding penetration depths are smaller suggesting similar yields.
Dividing the ion energy flux by an assumed average ion energy of 100 keV and multiplying by the x-ray yield per ion gives an x-ray surface emission rate of $1.5 \times 10^{6}$ photon/s-cm$^{2}$ for Europa and $1.2 \times 10^{6}$ photon/s-cm$^{2}$ for Io.  
If 10\% of these photons are O K-shell x-rays with energies around 525 eV, then the predicted power in O K-shell x-rays is $3.9 \times 10^{13}$ erg/s and $4.2 \times 10^{13}$ erg/s for Europa and Io, respectively.
The corresponding fluxes at the telescope are $8.1 \times 10^{-16}$ erg/s and $8.7 \times 10^{-16}$ erg/s for Europa and Io, respectively. 
The fluxes predicted by these simple estimates are within a factor of $\sim$three of the observed fluxes for Europa and Io.
Despite these low flux levels, further observations with the {\sl Chandra} and  XMM-Newton x-ray observatories may eventually permit the identification of lines from individual heavy elements.

\subsection{The Io plasma torus} \label{ss:distorus}

Turning now to the IPT, we estimate its x-ray flux (in O K-shell lines from O, O$^+$, and O$^{++}$) from fluorescence caused by the x-ray flux from the Sun.  
We use the results of model calculations for x-ray emission from the flaring active Sun (Peres {\sl et al.} 2000), normalized to the model x-ray luminosity in the energy band 0.1-10 keV, to determine the solar spectrum at Jupiter's orbit (the planet was 4.14 AU from the Earth and 4.96 AU from the Sun during the Nov 25-26, 1999, ACIS observations, and 4.12 AU from the Earth and 5.04 AU from the Sun during the Dec 18, 2000, HRC-I observations).
Photo-ionization cross-sections are from analytic fits to the results of calculations using the Hartree-Dirac-Slater method (Verner {\sl et al.} 1993).
For photo-ionization of oxygen ions at energies just above the K-edge, the cross-sections are of order few $\times$ 10$^{-19}$ cm$^{2}$, decreasing with increasing energy roughly as 1/E$^{3}$.
For the fluorescent yield we use the value for neutral oxygen (0.0083) (Krause 1979).
Galileo measured electron (Bagenal {\sl et al.} 1997) and ion (Crary {\sl et al.} 1998, Frank \& Paterson 2000) densities above 1000 cm$^{-3}$ in  some regions of the IPT.
Assuming the volume of a torus with semi-major axis equal to that of Io's orbit and diameter equal to Jupiter's equatorial diameter, and number densities in neutral oxygen and in ionized oxygen (mostly O$^+$) of 1000 cm$^{-3}$, we find a photon flux at the Earth of 2.6 $\times$ 10$^{-7}$ photons/s--cm$^{2}$ for the flaring active Sun.
Our spectral fits require photon fluxes in the apparent line near 570 eV of $\sim$3 $\times$ 10$^{-6}$ photons/s--cm$^{2}$, about an order of magnitude larger.
We conclude that the observed x-ray emission from the IPT is not due to fluorescence excited by solar x-rays.

Charge exchange processes have been invoked to explain Jupiter's x-ray aurora (Cravens {\sl et al.} 1995, Bhardwaj \& Gladstone 2000) and x-ray emission from comets (Cravens 1997, Dennerl 1999).
As energetic ions diffuse inward through the region of the Io plasma torus they are constantly subject to both electron stripping and charge exchange.
Electron stripping generally has a higher probability than charge exchange above some cross-over energy.  
The cross-over energy is different for the different charge states with higher charge states having a generally higher cross-over energy.
The cross-over energy for an OVII ion, above which electron stripping is favored, occurs above 11 MeV (Cravens 1997).
Ions below a few tens of keV rapidly charge exchange in the outer torus, become neutralized, and are lost from the Jupiter system (Mauk {\sl et al.} 1998).
These energetic neutrals have been recently observed by the Cassini spacecraft during its flyby of Jupiter (Mitchell {\sl et al.} 2001).
However, as the higher energy ions slowly diffuse radially inward through the plasma torus region, increasing numbers are driven to higher and higher charge states. 
In fact we argue that all oxygen ions with energies over 16 MeV are fully stripped by the time they diffuse radially inward to 6 Jupiter radii near Io's orbit (where they are observed).
Upon colliding with a torus neutral and undergoing charge exchange to O VII, such ions can produce an x-ray photon at energies near 570 eV.
From the Galileo Heavy Ion Counter experiment (Cohen {\sl et al.} 2000), we find  an integrated flux of oxygen ions above 16 MeV of 1.5 $\times$ 10$^{4}$ ions/s-cm$^{2}$ .
Combining this with an estimated charge exchange cross-section (Cravens {\sl et al.} 1995) of 10$^{-18}$ cm$^{2}$, an estimated neutral density of 30 cm$^{-3}$, and an estimated x-ray emission probability of 0.065, we obtain an x-ray volume production rate of 4.5 $\times$ 10$^{-13}$ photons/s-cm$^{3}$.
Taking the same IPT volume as assumed for our fluorescence estimate, we find 
a photon flux at Earth of $\sim$4 $\times$ 10$^{-10}$ photons/s-cm$^{2}$, approximately 4 orders of magnitude smaller than the required flux near 570 eV.
Similar considerations show that ion stripping by dust grains leading to x-ray emission also fails to explain the apparent line emission.

A large fraction of the IPT x-ray flux may result from bremsstrahlung radiation from non-thermal electrons. 
Above a few tens of eV, the electrons' energy distribution is well-described by a power law rather than a Maxwellian (Sittler \& Strobel 1987, Meyer-Vernet, Moncuquet, \& Hoang 1995). 
These non-thermal electrons play a critical role in setting the charge state of the Io torus, and the energy balance of the IPT. 
Estimates of cooling from electron-impact excited UV emissions and heating from Coulomb collisions with ions (Smith {\sl et al.} 1998) show the need for a large, additional source of energy. 
The nature of this energy source remains speculative, but it may result from wave-particle heating (Barbosa 1994) or inward diffusion of energetic ions and electrons from Jupiter's middle and outer magnetosphere (Smith {\sl et al.} 1998). 
Previous remote sensing observations have provided limited data on these non-thermal electrons, since UV and longer wavelength emissions are primarily sensitive to electrons with energies below a few eV.

Bremsstrahlung emission of soft X-rays are, on the other hand, produced
by the non-thermal electrons in the few hundred to few thousand eV range.
Based on the in-situ Ulysses observations, we assume the electrons are in a
kappa, or generalized Lorentzian, distribution with a temperature of 10 eV
and an index of $\kappa$ = 2.4 (Meyer-Vernet, Moncuquet, \& Hoang 1995). 
Using this distribution and the Bethe-Heitler cross-section for bremsstrahlung emission (with Sommerfeld-Elwert correction), we numerically calculate the source rate and spectrum. 
At a density of 2000 cm$^{-3}$, typical of the IPT, the source
rate is 6 $\times$ 10$^{-18}$ ergs/cm$^{3}$/s. 
Integrated over the entire IPT, the radiated x-ray power between 250 and 1000 eV is approximately 3 $\times$ 10$^{7}$ W, or roughly one third of the observed signal. 
The predicted shape of the bremsstrahlung spectrum is harder than the measured 1/E$^{6.8}$ spectrum, indicating that some of the flux near 250 eV is probably an extension of the FUV emission from the IPT (Hall {\sl et al.} 1994, Gladstone \& Hall 1998).

\section{Conclusions} \label{s:conclusions}

{\sl Chandra X-ray Observatory} observations have shown that Io and Europa, and probably Ganymede, emit soft x-rays with 0.25-2.0 keV luminosities $\sim$1-2 MW.  
This emission is likely to result from bombardment of their surfaces by energetic ($>$ 10 keV) H, O, and S ions from the region of the IPT.
The IPT itself emits soft x-rays with a 0.25-1.0 keV luminosity $\sim$0.1 GW.  
Most of this emission appears at the low end of the energy band, but an unresolved line or line complex is apparent in the spectrum at an energy consistent with oxygen.
The origin of the IPT x-ray emission is uncertain, but bremsstrahlung from non-thermal electrons may account for a significant fraction of the continuum x-rays.

Further x-ray observations and more detailed modelling are needed to probe more deeply into the origin and properties of the x-ray emission from the Galilean satellites and from the Io plasma torus, and their implications for the Jovian magnetosphere. \\

The authors thank L. Townsley for making CTI corrections to the ACIS data, and F. Bagenal for helpful communications.

\end{document}